\RequirePackage{fix-cm}
\documentclass[twocolumn]{svjour3}          
\smartqed  
\usepackage{graphicx}
\usepackage{braket}
\usepackage{dcolumn}
\usepackage{bm}
\usepackage{braket}
\usepackage{amsmath}
\usepackage{amssymb}
\usepackage{epstopdf}
\journalname{International Journal of Theoretical Physics}
\begin{document}

\title{Perfect quantum state transfer in Glauber-Fock cavity array
}


\author{Nilakantha Meher 
}


\institute{Indian Institute of Technology Kanpur\\
Uttar Pradesh 208016\\
India\\
              \email{nilakantha.meher6@gmail.com}           
}

\date{Received: date / Accepted: date}

\maketitle

\begin{abstract}
We study transfer of single photon in an one-dimensional finite Glauber-Fock cavity array whose coupling strengths satisfy a square root law. The evolved state in the array can be mapped to an upper truncated coherent state if the cavities are resonant. Appropriate choices of resonance frequencies provide perfect transfer of a photon between any two cavities in the array. Perfect transfer of the photon allows perfect quantum state transfer between the cavities. Our findings may help in realizing quantum communication and information  processing in photonic lattices. 
\end{abstract}

\section{Introduction}
Quantum state transfer is an essential task for realizing quantum communication \cite{Bose,Cirac,Meher,Bose2,Chapman,Chris2,Chris,Almeida,Zheng,LiJian}. This protocol requires transfer of an unknown qubit state between two nodes \cite{Reiserer,Northup,Zhong}. Many physical systems such as spin chain \cite{Bose,Chris,Chris2}, Josephson-junction array \cite{Lyakhov}, quantum dots \cite{Picoes,JiLi}, coupled cavities 
\cite{Meher,Almeida,YangLiu,Ogd}, photonic lattices \cite{Perez2,Perez3},  etc have been investigated to realize quantum state transfer. Possibility in precise control of resonance frequencies and coupling strengths of the array, coupled cavities provides a suitable physical system to realize controlled quantum state transfer \cite{Meher,Almeida,YangLiu,Yung,Neto,Hua}.\\

Cavities have been used to realize many interesting phenomena such as entanglement generation \cite{Leons,Liew,Browne,Miry}, quantum state preparation \cite{Yurk,Rojan,WeiWei,Yanhua}, localization-delocalization \cite{Meher2,Schmidt}, photon blockade \cite{Imamoglu,Birnbaum,Tang,Shen,Miranowicz} etc by properly tuning the resonance frequencies or including Kerr nonlinearity. In addition, coupling the cavities and suitably modifying the coupling strengths provide perfect transfer of photons between the cavities \cite{Meher,Almeida,Zhou,Qin}. \\ 


Such an array of cavities that has attracted a lot of attention is Glauber-Fock cavity array, whose coupling strengths satisfy $J_{k}=J\sqrt{k}$. This array is used for realizing classical analogue of coherent state and displaced Fock state \cite{Perez}, dynamic localization and self imaging \cite{Longhi}, transport and revival of squeezed light \cite{Rai}, Bloch-like revival \cite{Keil}, etc. In this article, we investigate perfect quantum state transfer between any two cavities in a Glauber-Fock cavity array. A single photon in the first cavity evolves to a state which can be mapped to a truncated coherent state. Suitable choices of resonance frequencies of the cavities allow perfect transfer of single photon between the cavities. This shows the possibility of transferring a qubit state of the form $\alpha\ket{0}+\beta\ket{1}$, where $\ket{0}$ is the vacuum state and $\ket{1}$ is the single photon state. \\

This article is organized as follows: In Sec. \ref{GFC}, we briefly overview Glauber-Fock cavity array. We show the possibility of perfect transfer of a photon between the cavities in the array in Sec. \ref{PerfectTransfer}. In Sec. \ref{StateTransfer}, we establish perfect transfer of a qubit state. Effect of dissipation on quantum state transfer is presented in Sec. \ref{ED}. Finally we summarized our results in Sec. \ref{Summary}.

\section{Glauber-fock cavity array}\label{GFC}
Consider a system of $N$ cavities described by the Hamiltonian 
\begin{equation}\label{Hamiltonian}
H=\sum_{k=1}^{N}\omega_k a_k^\dagger a_k +\sum_{k=1}^{N-1} J_{k}(a_k^\dagger a_{k+1}+a_k a_{k+1}^\dagger),
\end{equation}
where $\omega_k$ is the resonance frequency of the $k$th cavity. The operator $a_k (a_k^\dagger)$ is the annihilation (creation) operator for the $k$th cavity. The strength of coupling between $k$th cavity and $(k+1)$th cavity is $J_{k}$, which satisfies the square root law, \textit{i.e}, $J_{k}=J\sqrt{k}$. These coupling strengths can be achieved by adjusting the separation between the cavities with $J_{k}\sim e^{-\eta d_k}$ \cite{Perez}. Here $d_k$ is the distance between $k$th cavity and $(k+1)$th cavity,  and $\eta$ is a constant.\\

A key feature of $H$ given in Eqn. \ref{Hamiltonian} is that it conserves total number of quanta \textit{,i.e.,} $[H,\sum_k a_k^\dagger a_k]=0$. This signifies that there are invariant subspaces for the unitary dynamics generated by the Hamiltonian. Consider the cavities are resonant, then the Hamiltonian given in Eqn. \ref{Hamiltonian} in the interaction picture is
\begin{align}\label{interactionH}
H_{int}=\sum_{k=1}^{N-1} J_{k}(a_k^\dagger a_{k+1}+a_k a_{k+1}^\dagger).
\end{align}
We restrict the total number of quanta to be one. Then, the set $\ket{1_1,0_2,0_3,0_4...0_N},\ket{0_1,1_2,0_3,0_4,...0_N}$, ............, $\ket{0_1,0_2,0_3,...1_N}$ forms a basis, where $n_m$ inside the ket represents $n$ photons in $m$th cavity. For simplicity, the state $\ket{0_1,0_2,0_3,..,1_k,...0_N}$ which represents single photon in $k$th cavity and other cavities are in vacuum, is denoted as $\ket{k}\rangle$. Now, the interaction Hamiltonian given in Eqn. \ref{interactionH} can be written in a matrix form using the aforementioned basis set. The matrix form of $H_{int}$ is
\begin{equation}
H_{int}=\left[\begin{array}{cccccc}
0 & J & 0 & 0  & \dots\\
J & 0 & \sqrt{2}J & 0  & \dots\\
0 & \sqrt{2}J & 0 & \sqrt{3}J  & \dots\\
\vdots & \vdots &\vdots  &\ddots & \sqrt{N-1}J\\
0 & 0  & \dots & \sqrt{N-1}J & 0\\
\end{array}\right],
\end{equation}
which can be written in the form
\begin{align}\label{interaction}
H_{int}=J(A+A^\dagger),
\end{align}
where 
\begin{equation}\label{DefA}
A=\left[\begin{array}{cccccc}
0 & 1 & 0 & 0  & \dots\\
0 & 0 & \sqrt{2} & 0  & \dots\\
0 & 0 & 0 & \sqrt{3}  & \dots\\
\vdots & \vdots &\vdots &\ddots & \sqrt{N-1}\\
0 & 0  & \dots & 0  &0\\
\end{array}\right].
\end{equation}
It is to be noted that $A\ket{k}\rangle=\sqrt{k}\ket{k-1}\rangle$, which is analogous to annihilation operator.
Hence, the interaction Hamiltonian given in Eqn. \ref{interaction} can be mapped to a driven oscillator with the driving strength $J$. Now, a single photon in the first cavity evolves to
\begin{align}
\ket{\psi(t)}&=e^{-iH_{int}t}\ket{1_1,0_2,0_3,0_4...},\nonumber\\
&=e^{-iH_{int}t}\ket{1}\rangle,\nonumber\\
&=e^{-\frac{1}{2}J^2t^2[A,A^\dagger]}e^{-iJt A^\dagger}e^{-iJtA}\ket{1}\rangle,
\end{align}
where we have used the Baker-Hausdorf-Campbell theorem \cite{Gerry}. Using $[A,A^\dagger]=1$ and $A\ket{1}\rangle=0$,  the above equation can be written as
\begin{align}
\ket{\psi(t)}&=e^{-\frac{1}{2}J^2t^2}\sum_{j} \frac{(-iJtA^\dagger)^j}{j!}\ket{1}\rangle.
\end{align}
Now, using the action of $A^\dagger$, \textit{i.e.,} $A^\dagger \ket{k}\rangle=\sqrt{k+1}\ket{k+1}\rangle$, the evolved state becomes
\begin{align}\label{EvolvedState}
\ket{\psi(t)}&=N_c\sum_{k=1}^N \frac{(-iJt)^{(k-1)}}{\sqrt{(k-1)!}}\ket{k}\rangle,
\end{align}
where
\begin{align}
N_c=\frac{e^{-(Jt)^2/2}}{\sqrt{\left[e^{-(Jt)^2}\sum_{j=0}^{N-1}\frac{(Jt)^{2j}}{j!}\right]}},
\end{align}
is the normalization constant.
Using the definition of incomplete gamma function \cite{Gradshteyn} 
\begin{align}
\gamma(N,x)=(N-1)!\left[1-e^{-x}\sum_{j=0}^{N-1}\frac{x^j}{j!}\right],
\end{align}
normalization constant becomes
\begin{align}
N_c=\frac{e^{-(Jt)^2/2}}{\sqrt{1-\frac{\gamma(N,(Jt)^2)}{(N-1)!}}}.
\end{align}
The evolved state given in Eqn. \ref{EvolvedState} is equivalent to an upper truncated coherent state  \cite{Le,MiranTanas,Meher3}.
In the limit of $N\rightarrow \infty$, the state is analogous to a coherent state $\ket{\alpha}$.\\

\begin{figure}[h]
\centering
\includegraphics[height=7cm,width=9.2cm]{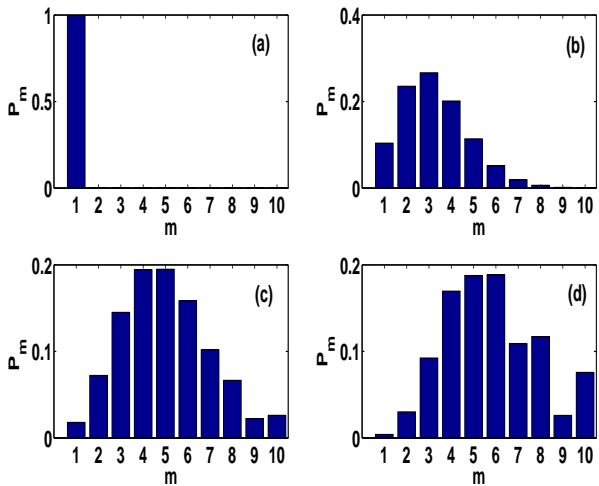}
\caption{ Probability distribution $P_m$ at $(a)\omega t=0,(b)\omega t=30,(c)\omega t=40$ and $(d) \omega t=84$. We set $J/\omega=0.05, N=10$.  }
\label{SinglePhotonResonantCase}
\end{figure}

Probability of detecting the single photon in $m$th cavity is 
\begin{align}\label{ProbabilityDistribution}
P_m=|\langle\langle m\ket{\psi(t)}|^2=\left|N_c\frac{(-iJt)^{(m-1)}}{\sqrt{(m-1)!}}\right|^2.
\end{align} 
Fig. \ref{SinglePhotonResonantCase} shows the probability distribution $P_m$ at various values of $\omega t$. This probability distribution is analogous to the photon number distributions in truncated coherent states with an appropriate amplitude.\\

In the context of perfect transfer, it can be infer from the figure that complete transfer of the photon to a specific cavity with unit probability is not possible. Hence controlling the photon transfer is not achievable in resonant cavities for this choice of coupling strength.
\section{Perfect transfer of a photon}\label{PerfectTransfer}
\begin{figure}[h]
\centering
\includegraphics[height=6cm,width=9cm]{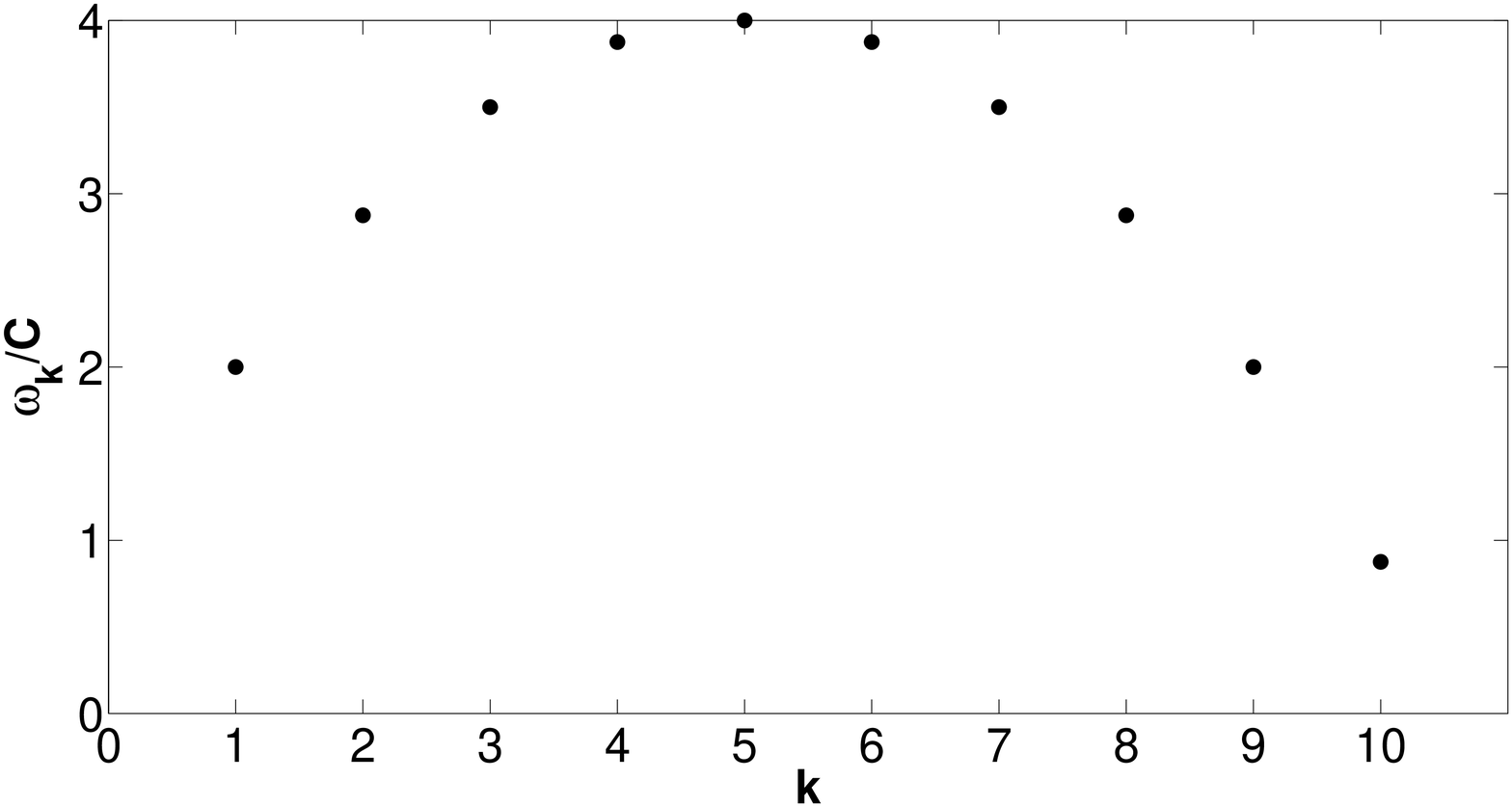}
\caption{ Resonance frequencies $\omega_k/C$ as a function of $k$. We choose $m=3$ and $n=7$ with $N=10$. Note that $\omega_3=\omega_7$. }
\label{Wkvsk}
\end{figure}

As seen in previous section, perfect transfer of a photon between any two cavities in the array is not possible if the cavities are resonant. A feature that can be used for controlling the transfer is to suitably choosing the resonance frequencies. Perfect transfer between any two cavities is already been achieved in an array by suitably modifying the resonance frequencies  where the coupling strengths are symmetric about the centre \cite{Meher}. A parabolic form of resonance frequencies are used for controlling the transfer \cite{Meher}. In this article, we  choose the resonance frequencies as 
\begin{align}\label{SwitchingCond}
\omega_k=C+\left((k-1)-\frac{(k-1)^2}{(m+n-2)}\right),
\end{align}
which are also parabolic (inverted). This choice of resonance frequencies guarantees $\omega_m=\omega_n$, \textit{i.e.,} 
\begin{align}
\langle \bra{m}H\ket{m}\rangle=\langle \bra{n}H\ket{n}\rangle,
\end{align}
for any $m$ and $n$. This is the basic requirement for transferring a photon from $m$th cavity to $n$th cavity. These resonance frequencies are shown in Fig. \ref{Wkvsk}, where we choose $m=3$ and $n=7$. The frequencies that is given in Eqn. \ref{SwitchingCond} can be realized in photonic crystal cavities where precisely controlling the resonance frequencies is possible by various techniques such as nanofluidic tuning \cite{Vignolini2}, nanomechanical tuning \cite{Hopman}, photochromic tuning \cite{Cai}, to mention a few.\\

Using degenerate perturbation theory, it can be seen that the states
\begin{align}
\ket{\pm}\rangle&=\frac{1}{\sqrt{2}}(\ket{m}\rangle\pm\ket{n}\rangle)+\mathcal{O}(J/\Delta_{m,m+1})\nonumber\\
&+\mathcal{O}(J/\Delta_{m,m-1})+\mathcal{O}(J/\Delta_{n,n+1})+\mathcal{O}(J/\Delta_{n,n-1}),
\end{align}
become the approximate eigenstates of $H$ given in Eqn. \ref{Hamiltonian}. Here, $\Delta_{x,y}=\omega_y-\omega_x$. The corrections that is given in the above equation are small for smaller $J$.\\

If the initial state is $\ket{m}\rangle$, \textit{i.e.,} a single photon in the $m$th cavity, then the state at later time is
\begin{align}\label{Evolvedstate2}
\ket{\psi(t)}&=e^{-iHt}\ket{m}\rangle,\nonumber\\
&\approx\frac{1}{\sqrt{2}} e^{-iHt}[\ket{+}\rangle+\ket{-}\rangle]+\mathcal{O}(J/\Delta_{m,m+1})\nonumber\\
&+\mathcal{O}(J/\Delta_{m,m-1})+\mathcal{O}(J/\Delta_{n,n+1})+\mathcal{O}(J/\Delta_{n,n-1}),\nonumber\\
&\approx e^{-i\lambda t}\left[ \cos\theta t\ket{m}\rangle-i\sin\theta t\ket{n}\rangle \right]+\mathcal{O}(J/\Delta_{m,m+1})\nonumber\\
&+\mathcal{O}(J/\Delta_{m,m-1})+\mathcal{O}(J/\Delta_{n,n+1})+\mathcal{O}(J/\Delta_{n,n-1}),
\end{align}
with $\theta=(\lambda_+-\lambda_-)/2$ and $\lambda=(\lambda_++\lambda_-)/2$. Here $\lambda_{\pm}$ are the eigenvalues of $H$ correspond to the approximate eigenstates $\ket{\pm}\rangle$. If $J$ is very small, the above equation can be written as
\begin{align}
\ket{\psi(t)}\approx e^{-i\lambda t}\left[ \cos\theta t\ket{m}\rangle-i\sin\theta t\ket{n}\rangle \right].
\end{align} 
Note that the photon is exchanged periodically between $m$th cavity and $n$th cavity. The probability of transferring a photon from $m$th cavity to $n$th cavity is 
\begin{align}
P_{mn}\approx\sin^2\theta t.
\end{align} 
At time $t=\pi/(\lambda_+-\lambda_-)$, photon is completely transferred from $m$th cavity to $n$th cavity. This is the minimum time to transfer a photon from $m$th cavity to $n$th cavity. Thus the condition given in Eqn. \ref{SwitchingCond} ensures perfect transfer of a photon between any two cavities in the array. \\

\begin{figure}[h]
\centering
\includegraphics[height=6cm,width=9cm]{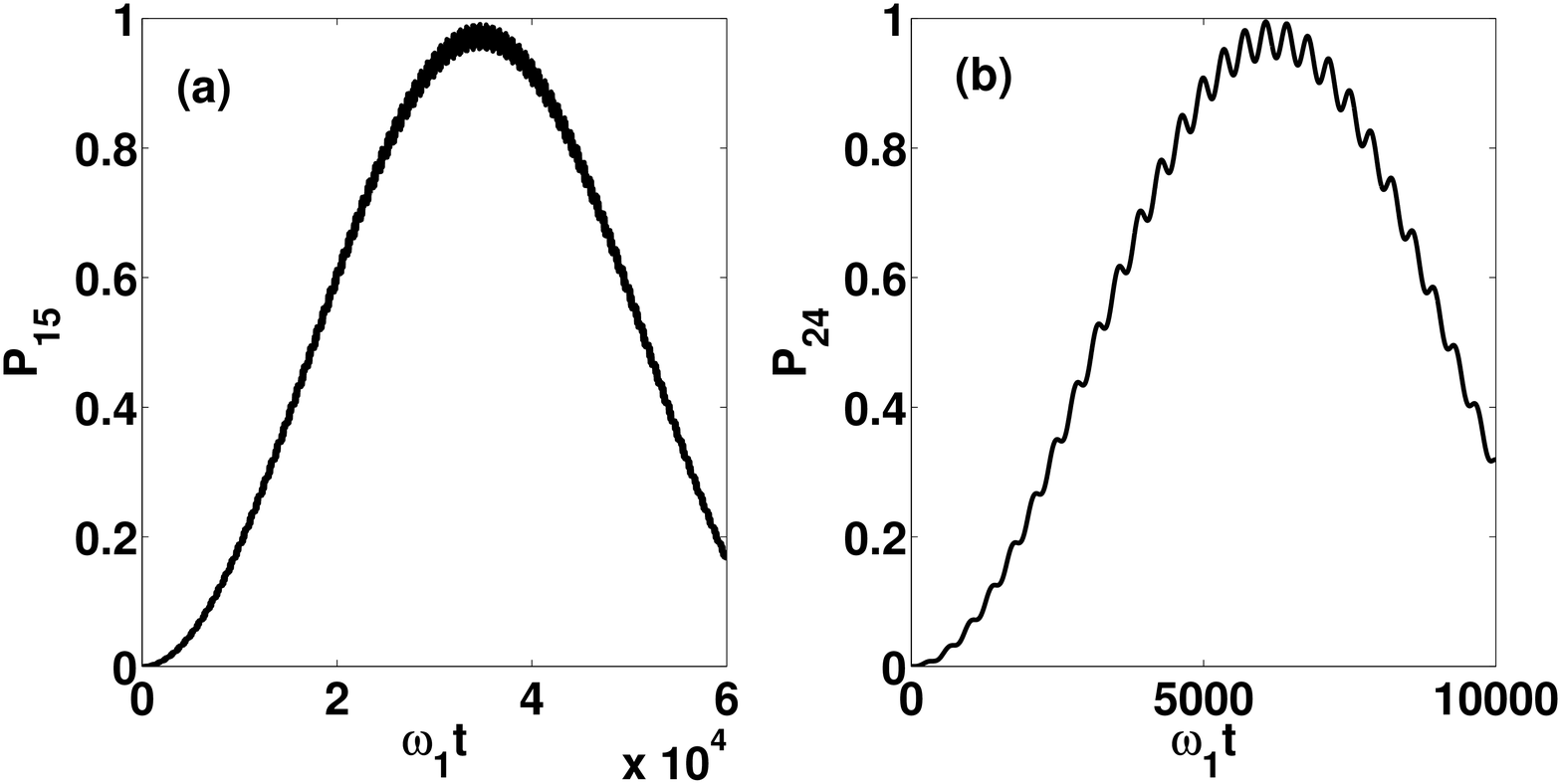}
\caption{ Probability of single photon transfer from $m$th cavity to $n$th cavity $(P_{mn})$ as a function of $\omega_1t$. Here we set $J/\omega_1=0.0013$. All the system parameters are in the unit of $\omega_1$. Resonance frequencies are satisfying the relation given in Eqn. \ref{SwitchingCond}.  }
\label{P15P24}
\end{figure}

Fig. \ref{P15P24}$(a)$ shows the probability of transferring a single photon from first cavity $(m=1)$ to fifth cavity $(n=5)$ and Fig. \ref{P15P24}$(b)$ shows the transfer probability from second cavity $(m=2)$ to fourth cavity $(n=4)$. Total number of cavities in the array is six. It can be seen in figure that there is perfect transfer of a photon from $m$th cavity to $n$th cavity. \\ 

An interesting aspect of this transfer is that the other cavities in the array are not populated significantly during the evolution. This comes from the fact that the states other than $\ket{m}\rangle$ and $\ket{n}\rangle$ do not contribute appreciably to $\ket{\psi(t)}$ given in Eqn. \ref{Evolvedstate2}.
\section{Quantum state transfer}\label{StateTransfer}
Perfect transfer of a photon between any two cavities in the array is possible if the resonance frequencies of the array satisfy the relation given in Eqn. \ref{SwitchingCond}. In this section, we establish perfect transfer of a quantum state of the form $\alpha\ket{0}+\beta\ket{1}$, which is the superposition of vacuum state and a single photon state.\\

Consider the interaction term in $H$ to be of the form
\begin{align}
H_{int}=\sum_{k=1}^{N-1} J_{k+1}(e^{i\eta} a_k^\dagger a_{k+1}+ e^{-i\eta} a_k a_{k+1}^\dagger),
\end{align}
where the coupling strengths are assume to be complex. Complex coupling strengths can be realized experimentally using wave-guide delay line \cite{Hafezi}. 
Now, the approximate eigenvectors of $H$ are 
\begin{align}
\ket{\pm}\rangle\approx\frac{1}{\sqrt{2}}\left[\ket{m}\rangle\pm e^{i(n-m)\eta}\ket{n}\rangle \right].
\end{align}
Consider the initial state of the cavity array to be $\alpha\ket{\text{vac}}\rangle+\beta\ket{m}\rangle$, which corresponds to the $m$th cavity in the superposition $\alpha\ket{0}+\beta\ket{1}$ and the other cavities are in their respective vacuua.  The target state is $\ket{\Psi}=\alpha\ket{\text{vac}}\rangle+\beta\ket{n}\rangle$, which corresponds to the $n$th cavity in the superposition $\alpha\ket{0}+\beta\ket{1}$ and the other cavities are in vacuum. \\

Now the initial state evolves as
\begin{align}
\ket{\psi(t)}=\alpha\ket{\text{vac}}\rangle &+\beta e^{-i\lambda t}[\cos\theta t\ket{m}\rangle \nonumber\\
&-ie^{-i(n-m)\eta}\sin\theta t  \ket{n}\rangle].
\end{align}
Fidelity for the quantum state transfer from $m$th cavity to $n$th cavity is
\begin{align}
F_{mn}=&\left|\langle\Psi|\psi(t)\rangle\right|^2,\nonumber\\
\approx&\left| |\alpha|^2-i|\beta|^2 e^{-i\lambda t}e^{-i(n-m)\eta}\sin\theta t \right|^2,
\end{align}
where $\lambda=(\lambda_++\lambda_-)/2$.
If $\eta=\frac{1}{(m-n)}(\frac{\pi}{2}-\frac{\lambda\pi}{2\theta})$, the fidelity becomes nearly unity at $\theta t=\pi/2$. In other words, the state of the $n$th cavity is in the superposition $\alpha\ket{0}+\beta\ket{1}$ and rest of the cavities are in vacuum.\\

\begin{figure}[h]
\centering
\includegraphics[height=7cm,width=9.2cm]{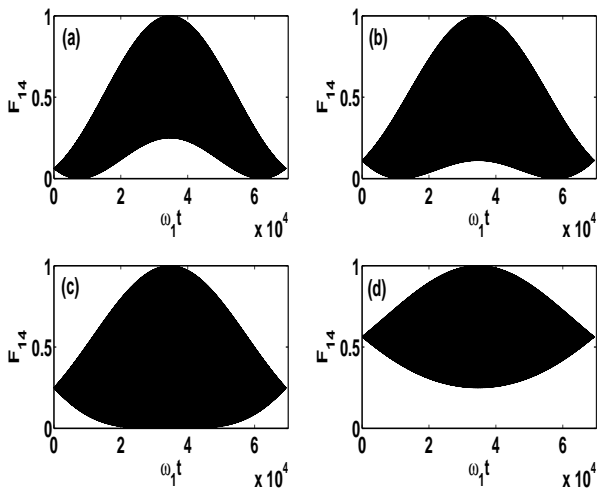}
\caption{Fidelity of quantum state transfer from 1st cavity to 4th cavity $(F_{14})$ where $(a)\alpha=1/2, \beta=\sqrt{3}/2$, $(b)\alpha=1/\sqrt{3}, \beta=i\sqrt{2/3}$, $(c)\alpha=1/\sqrt{2}, \beta=1/\sqrt{2}$ and $(d)\alpha=\sqrt{3}/2, \beta=1/2$. Here we set $J/\omega_1=0.0013$. Resonance frequencies are satisfying the relation given in Eqn. \ref{SwitchingCond}. All the system parameters are in the unit of $\omega_1$. }
\label{F14forvariousstates}
\end{figure}
Fig. \ref{F14forvariousstates} shows the fidelity $F_{14}$, \textit{i.e.}, fidelity of quantum state transfer from first cavity to fourth cavity as a function of $\omega_1 t$ for various choices of $\alpha$ and $\beta$. It is to be noted that the fidelity is unity for all choices of $\alpha$ and $\beta$. Hence,  the condition given in Eqn. \ref{SwitchingCond} ensures perfect transfer of a quantum state between any two cavities in the array.
\section{Effect of dissipation}\label{ED}
An ideal system is characterized by complete isolation of the system from environment. However, dissipation is unavoidable. Effect of dissipation on quantum state transfer fidelity is studied by analyzing the master equation.  Master equation that includes dissipation in the cavity array is \cite{Carmichael}
\begin{align}\label{Master}
\frac{\partial \rho}{\partial t}=-i[H,\rho]+\frac{\gamma}{2}\sum_{k=1}^N \mathcal{L}(a_i)\rho,
\end{align}
where
\begin{align*}
\mathcal{L}(o)\rho=(2o\rho o^\dagger-o^\dagger o \rho-\rho o^\dagger o),\\
\end{align*}
is the Lindblad superoperator \cite{Lindblad}. For a single cavity, master equation can be solved analytically by super-operator method \cite{Arevalo,Mufti}. However, for coupled cavities, the number of equations of motion depends on the dimension of density matrix and hence, an analytical solution for master equation is difficult to derive. Here, we solve the master equation numerically.   The equations of motion for the density matrix elements $\rho_{m,n}$ are
\begin{align}
\frac{\partial\rho_{m,n}}{\partial t}&=\frac{\partial}{\partial t}\langle \bra{m}\rho\ket{n}\rangle, \nonumber\\
&=-i[(\omega_m-\omega_n)\rho_{m,n}+J_m(\rho_{m+1,n}+\rho_{m-1,n})\nonumber\\
&+J_n(\rho_{m,n+1}+\rho_{m,n-1})]+\gamma(2\rho_{m+1,n+1}-\rho_{m,n}).
\end{align}
By considering the initial state $\ket{\psi_{in}}=\alpha\ket{\text{vac}}\rangle+\beta\ket{m}\rangle$ and numerically solving the above equations of motion one can obtain the density matrix at various instances of time.\\
\begin{figure}[h]
\centering
\includegraphics[height=7cm,width=9cm]{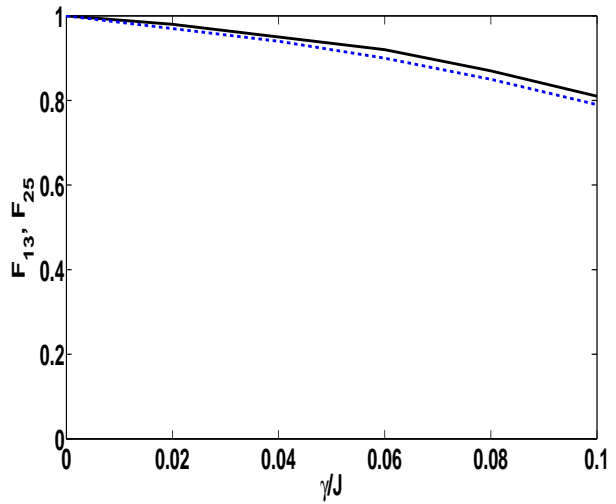}
\caption{ Average fidelities of quantum state transfer $F_{13}$ (continuous) and $F_{25}$ (dashed) as a function of $\gamma/J$. All the resonance frequencies satisfy the relation given in Eqn. \ref{SwitchingCond}. }
\label{F13F25VsGamma}
\end{figure}

Fidelity for the quantum state transfer in the presence of dissipation can be calculated as
\begin{align}
F_{mn}=\left[\text{Tr}\left(\sqrt{\sqrt{\sigma}\rho\sqrt{\sigma}}\right)\right]^2.
\end{align}
Here $\sigma=\ket{\Psi}\bra{\Psi}$, where $\ket{\Psi}=\alpha\ket{\text{vac}}\rangle+\beta\ket{n}\rangle$.\\ 

Fig. \ref{F13F25VsGamma} shows average of fidelities of quantum state transfer between the cavities as a function of $\gamma/J$. Here the average is taken over ensemble of initial states, \textit{i.e.}, $\alpha$ and $\beta$ are chosen randomly. In the absence of dissipation, the fidelities are nearly unity. Average fidelities decrease due to photon loss from the array.
\section{Summary}\label{Summary}
Quantum state transfer is essential for future quantum communication. We studied single photon transfer and quantum state transfer in Glauber-Fock cavity array whose coupling strengths satisfy square root law. If the cavities are resonant, the evolved state in the array can be mapped to an upper truncated coherent state. By suitably modifying resonance frequencies of the cavities, we achieve perfect transfer of single photon between any two cavities in the array. Transfer of a photon allows us to transfer of a qubit state between the cavities. High fidelity of quantum state transfer is achieved even in the presence of dissipation. These results are useful in the context of quantum information processing and communication. The ideas presented here are also applicable for spin chains if the coupling strengths are suitably modified.
\begin{acknowledgements}
NM acknowledges Indian Institute of Technology Kanpur for postdoctoral fellowship.
\end{acknowledgements}

%
 \section*{Conflict of interest}
 The authors declare that they have no conflict of interest.



\end{document}